\documentclass[]{article}

\usepackage{epsfig}
\usepackage{graphicx}
\usepackage{times,epsfig,makeidx,amsfonts,amsmath}
\usepackage{lscape} 
\usepackage{color}
\usepackage{caption}
\usepackage{subcaption}
\usepackage{authblk}

\def\d{{\rm d}}

\def\half{\hbox{$1\over2$}}

\begin{document}

\title{Modeling with a Large Class of Unimodal Multivariate Distributions}

\author[1]{Marina Silva Paez}
\author[2]{Stephen G. Walker}
\small
\affil[1]{Instituto de Matem\'atica, Universidade Federal do Rio de Janeiro, Brazil}
\affil[2]{Department of Statistics and Data Sciences, University of Texas at Austin, U.S.A}
\normalsize
\renewcommand\Authands{ and }
\date{}
\maketitle

\begin{abstract}
In this paper we introduce a new class of multivariate unimodal distributions, motivated by Khintchine's representation. We start by proposing a univariate model, whose support covers all the unimodal distributions on the real line. The proposed class of unimodal distributions can be naturally extended to higher dimensions, by using the multivariate Gaussian copula. Under both univariate and multivariate settings, we provide MCMC algorithms to perform inference about the model parameters and predictive densities. The methodology is illustrated with univariate and bivariate examples, and with variables taken from a real data-set. 
\end{abstract}

\bigskip


\section{Introduction} The clustering of data into groups is one of the current major topics under research. It is fundamental in many statistical problems; most notably in machine learning, see, for example, \cite{TehJor10}. 

The traditional approach is to model the data via a mixture model, see \cite{Titterington85} for finite mixture distributions, and, more recently, \cite{frauhwirth06}. The idea, which carries through to infinite mixture models, is that each component in the mixture is represented by a parametric density function, typically from the same family, which is usually the normal distribution. This assumption, often made for convenience, supposes each cluster or group can be adequately modeled via a normal distribution. For if a group requires two such normals, then it is deemed there are in fact two groups.  Yet two normals could be needed even if there is one group and it happens to be skewed, for example.

On the other hand, if we start by thinking about what a cluster could be represented by in terms of probability density functions, then a unimodal density is the most obvious choice. Quite simply, with lack of further knowledge, i.e. only observing the data, a bimodal density would indicate two clusters. This was the motivation behind \cite{rodriguezwalker14}, which relies heavily on the unimodal density models being defined on the real line, for which there are no obvious extensions to the multivariate setting.  

There are representations of unimodal density functions on the real line (e.g. \cite{Khint38} and \cite{Feller71}) and it is not difficult to model such a density adequately. See, for example, \cite{brunner89}, \cite{quintana09} and \cite{rodriguezwalker14}. Clearly, the aim is to provide large classes of unimodal densities and hence infinite dimensional models are often used. 

However, while there is an abundance of literature on unimodal density function modeling on the real line, there is a noticeable lack of work in two and higher dimensions. The reasons are quite straightforward; first there is a lack of a representation for unimodal distributions outside of the real line and, second, the current approaches to modeling unimodal densities on the real line do not naturally extend to higher dimensions. 

The aim in this paper is to introduce and demonstrate a class of unimodal densities for a multivariate setting. While marginal density functions will be modeled using the \cite{Khint38} representation on the real line, the dependence structure will be developed using a Gaussian copula model. To elaborate, using an alternative form of representation of unimodal densities on the real line,  we can write a unimodal density as
\begin{equation}
f(y)=\int_0^1 f(y|x)\,\d x.\label{unim}
\end{equation}
This then naturally extends to a bivariate setting with marginals of  the form (\ref{unim})  via the use of a copula density function $c(x_1,x_2)$;
$$f(y_1,y_2)=\int_0^1\int_0^1 f_1(y_1|x_1)\,f_2(y_2|x_2)\,c(x_1,x_2)\,\d x_1\,\d x_2.$$
Using a Gaussian copula, which has the ability to model pairwise dependence, we can then easily proceed upwards to the general multivariate unimodal density. 

The layout of the paper is as follows: In section 2 we provide the key representation of unimodal densities on the real line which we adapt in order to define continuous densities. Otherwise there is an issue of continuity at the mode. With a full representation of unimodal densities, we need a novel MCMC algorithm, particularly, and perhaps strangely, to include a non-zero mode. Section 3 then deals with the multivariate setting. There are peculiarities to the higher dimension case with the MCMC and again it is the location of the mode parameter which needs special attention. Section 4 provides a real data analysis.

\section{Models for unimodal densities}

Our aim in this paper is to start with the representation of \cite{Khint38} for unimodal densities on the real line and extend it to higher dimensions in a natural way, using the multivariate Gaussian copula. The representation of \cite{Khint38} is given by
$Y=X\,Z$, where $X$ is uniform on $[0,1]$ and $Z$, independent of $X$, has any distribution. Then $Y$ has a unimodal distribution with mode at $0$.  To see this let us consider the situation of $y>0$. So
$$P(Y\leq y)=\int_0^1 P(Z\leq y/x)\,\d x$$
and hence
$$f(y)=\int_0^1 x^{-1}\,g(y/x)\,\d x,$$
where $g$ is used to represent the density function of $Z$.
To see more clearly the unimodality at $0$, let us use the transform $s=y/x$, so
\begin{equation}
f(y)=\int_y^\infty s^{-1}\,g(s)\,\d s,\label{oldf}
\end{equation}
which is maximized at $y=0$.

This representation has been widely used and is usually presented as a mixture of uniform distribution; i.e.
$$f(y)=\int s^{-1}{\bf 1}(0<y<s)\,\d G(s).$$
The prior for $G$, in a Bayesian nonparametric setting, is usually taken as a Dirichlet process (\cite{Ferg73}). To denote this we write $G \sim D(M,G_0)$, where $M > 0$ is a scale parameter, and $G_0$ a distribution function. In particular, $E(G)=G_0$.

Now \cite{brunner89} and \cite{quintana09}, among others, have worked with this specification and extended to the real line with arbitrary mode by using $U(y|\kappa-s,\kappa+s)$ for some $\kappa\geq 0$. This specification leads to a symmetric  density. 

Furthermore,  \cite{rodriguezwalker14} develop the model to allow for asymmetry by adopting the idea of \cite{fernandezsteel98}; incorporating an asymmetry parameter $\lambda$ in the uniform kernel,
\begin{eqnarray}
f(y|\lambda, \kappa, G) &=& \int U\left(y|\kappa- s e^{-\lambda}, \kappa + s e^{-\lambda}\right) \,\d G(s).  \label{RW2012}
\end{eqnarray}
So (\ref{RW2012}) defines a unimodal density determined by $(\lambda, \kappa, G)$, where $\kappa$ is the location parameter, $\lambda$ defines the asymmetry, and the distribution $G$ defines characteristics such as variance, kurtosis, tails and higher moments.  With this representation, the support of the model proposed by \cite{rodriguezwalker14} includes all symmetric unimodal densities and a large class of asymmetric ones.

However, when extending to the real line, and using a natural density for $g$ such as the normal, we encounter the situation of $f(0)=\infty$. In fact, on the real line, $g$ needs to be somewhat unusual to ensure that $f(0)\ne \infty$.  See the form of density in (\ref{oldf}). 

To resolve this problem, we instead work with $Y=X/Z$; so again looking at $y>0$, we have
$$P(Y\leq y)=\int_0^1 P(Z\geq x/y)\,\d x.$$
Hence,
\begin{equation}
f(y)=\int_0^1 (x/y^2)\,g(x/y)\,\d x\label{origuni}
\end{equation}
and using the transform $s=x/y$ we obtain
\begin{equation}
f(y)=\int_0^{1/y} s\,g(s)\,\d s.\label{posuni}
\end{equation}
Thus we see the mode is again at $y=0$ and $f(0)$ will be finite subject to the more reasonable assumption that $sg(s)$ is integrable at $0$, rather than the more unreasonable assumption that $s^{-1}g(s)$ is integrable at $0$ in the former setting.

\subsection{A class of unimodal densities} \label{unimoddens}

Here we further investigate the choice of $Z$ being a normal distribution with the representation $Y=X/Z$. We can easily extend (\ref{posuni}) to the whole real line; resulting in
\begin{equation}
f(y)={\bf 1}(y>0)\,\int_0^{1/y}\,s\,g(s)\,\d s+{\bf 1}(y<0)\,\int_{1/y}^0 |s|\,g(s)\,\,\d s.\label{realuni}
\end{equation}
For a mode at $\kappa\ne 0$ we simply exchange $y$ for $y-\kappa$.
If the mode is at $0$, then for $\lim_{y \downarrow 0} f(y) =\lim_{y \uparrow 0} f(y)$, we need 
$$
\int_0^\infty s g(s) \d s = \int_\infty^0 |s| g(s) \d s. 
$$
Our aim is to take $g$ as a large class of density functions on the real line and the full class is given by mixtures of normal distributions. Thus, it is expedient to compute (\ref{realuni}) for $g(s)$ normal with mean $\mu$ and variance $\sigma^2$. That is, for $\xi>0$
$$f(\xi)=\int_0^\xi s\,N(\d s|\mu,\sigma^2),$$
which, after some algebra, results in
$$f(\xi)=\mu\left[\Phi\left(\frac{\xi-\mu}{\sigma}\right)-\Phi\left(\frac{-\mu}{\sigma}\right)\right]
+\sigma\left[\phi\left(\frac{-\mu}{\sigma}\right)-\phi\left(\frac{\xi-\mu}{\sigma}\right)\right],$$
where $\phi$ and $\Phi$ are the pdf and cdf, respectively,  for the standard normal distribution. 
With $\xi<0$ we simply need to rearrange the signs, so
\begin{equation}
f(\xi)=\left|\mu\left[\Phi\left(\frac{\xi-\mu}{\sigma}\right)-\Phi\left(\frac{-\mu}{\sigma}\right)\right]
+\sigma\left[\phi\left(\frac{-\mu}{\sigma}\right)-\phi\left(\frac{\xi-\mu}{\sigma}\right)\right] \right|.\label{moduni}
\end{equation}
Hence, this $f(\xi)$ holds for all $\xi \in (-\infty, +\infty)$.

To move the mode to $\kappa$, rather than $0$, write $f(y|\mu,\sigma,\kappa)=f(1/(y-\kappa))$ with $f(\cdot)$ given by (\ref{moduni}).  We could also maintain the representation (\ref{origuni}) in which case we can write
$$
f(y|\mu,\sigma,\kappa)=\frac{1}{\sigma\sqrt{2\pi}}\int_0^1 \left\{ x/(y-\kappa) \right\}^2\,\exp\left\{-\half\left(\frac{x/(y-\kappa)-\mu}{\sigma}\right)^2\right\}\,\d x.
$$
The usefulness of this is that it provides a latent joint density function which will be helpful when it comes to model estimation, namely,
\begin{eqnarray}
f(y,x|\mu,\sigma,\kappa)\propto \left\{ x/(y-\kappa) \right\}^2\,\exp\left\{-\half\left(\frac{x/(y-\kappa)-\mu}{\sigma}\right)^2\right\}. \label{withx}
\end{eqnarray}
For the forthcoming mixture model, where we consider mixtures of normals for $Z$, we find it convenient for identifiability concerns, to use the parametrization $(\mu,c)$, where $c=(\mu/\sigma)^2$.

\subsubsection{The mixture model}

The model proposed for the class of unimodal density functions on ${\rm I\!R}$ is a mixture of the unimodal densities with mode at $\kappa$ proposed in section \ref{unimoddens}, i.e. (\ref{withx}). This model can we written as
\begin{eqnarray}
f(y) = \sum_{j=1}^\infty w_j f(y|\mu_j,c, \kappa). \label{inftysum}
\end{eqnarray}
The mode is therefore at $\kappa$ and we allow parameter $c$, which is a transformation of the coefficient of variation, to remain fixed across the normal components. This is analogous to keeping the variance terms fixed, for identifiability concerns, in a mixture of normal model. The support of the model is not diminished by doing this.

\begin{figure}
	\centering
\includegraphics[width=10cm,height=5cm]{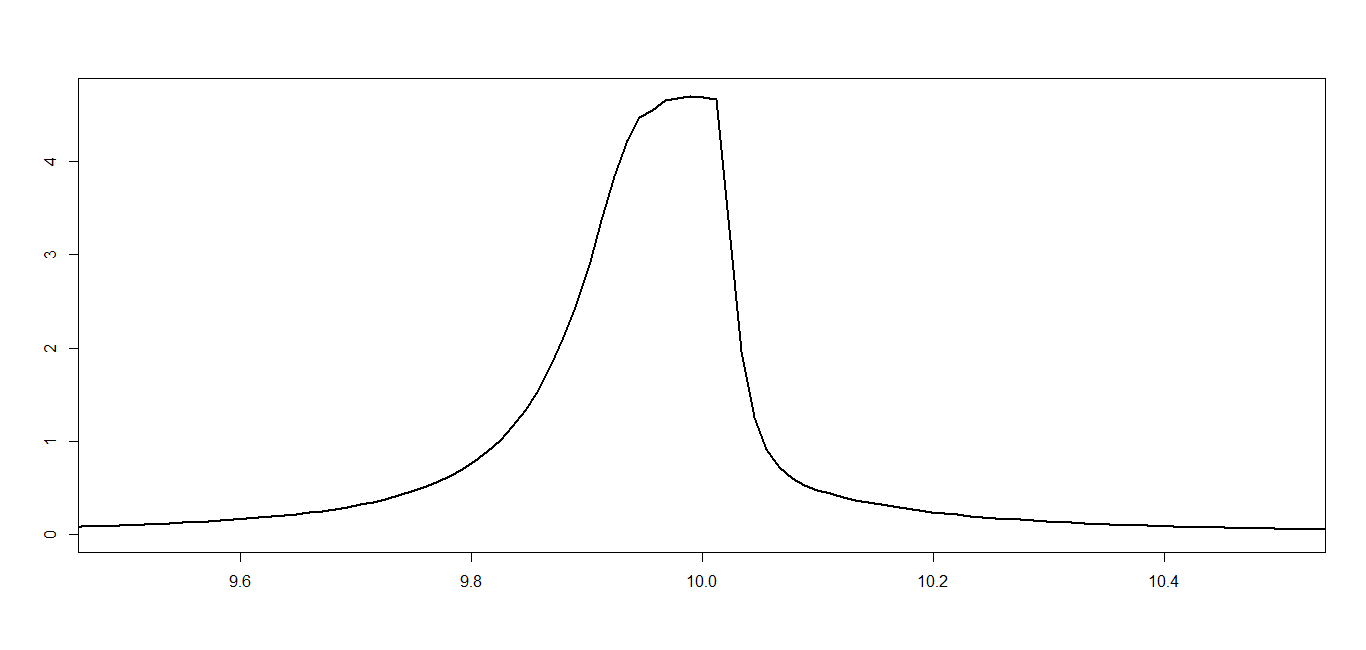} 
\caption{Unimodal density function (\ref{withx}) with a mixture of two components, $\kappa = 10$, $\mu = (-5,20)$, $w = (0.8,0.2)$.}
\label{density}
\end{figure}

Therefore, we can write
\begin{equation}
f(y)=\int f(y|\mu,c,\kappa)\,\d G(\mu)\label{mixmod}
\end{equation}
where $G$ is a discrete distribution with mass $w_j$ at the point $\mu_j$. Hence, in a Bayesian \\ context, the \hspace{0.01 mm} parameters \hspace{0.01 mm} to \hspace{0.01cm} be \hspace{0.01cm} estimated \hspace{0.01cm} and \hspace{0.01cm} first \hspace{0.01cm} assigned \hspace{0.01cm} prior \hspace{0.01cm} distributions are \\ $((\mu_j,w_j),c,\kappa)$, with the constraints being that the weights sum to 1, $c>0$, and the $(\mu_j)$ are distinct. As an illustration, Figure \ref{density} shows the density function of a mixture of two components with $\kappa = 10$, $c=1$, $\mu = (-5,20)$ and $w = (0.8,0.2)$. 

The claim is that model \ref{mixmod} can be arbitrarily close to any unimodal density on the real line. The key is writing $y=X/Z$ rather than $y = XZ$, and with the former we can employ the full support abilities of a mixture of normals without the problem of a discontinuity at the mode.

\subsection{Prior specifications and model estimation} \label{priors}

\subsubsection{Prior specifications}

To complete the model described in the previous section, we now set the prior distributions for the unknown model parameters. Specific values are assigned in the illustration section. 

\begin{itemize}

\item The prior distribution for the parameters $(\mu_j)$  are normal with zero mean and variance $\sigma_\mu^2$; while the prior for $\kappa$ will also be normal with zero mean and variance $\sigma_\kappa^2$. The prior for $c$ will be gamma with parameters $(\alpha_c,\beta_c)$.

\item The prior distribution for the weights $(w_j)$ has a stick-breaking construction (\cite{sethuraman94}), given by
$$w_1 = v_1 \quad\mbox{and}\quad w_j = v_j \prod_{l < j}(1 - v_l)\quad\mbox{for   }j>1,$$
with the $(v_j)$ being i.i.d.  from a beta distribution with parameters $1$ and $M$. We assume a gamma, $ga(\alpha_M,\beta_M)$, as the hyper-prior for the parameter $M$.
\end{itemize}

As a consequence, the prior for $G$ in (\ref{mixmod}) is a Dirichlet process with mean distribution
$N(0,\sigma_\mu^2)$ and scale parameter $M$.

Before detailing the MCMC algorithm for this model, we will now introduce the key latent variables which facilitate an ``as easy as possible'' implementable algorithm.
First, to deal with the infinite sum of elements in (\ref{inftysum}), we introduce a variable $u$, such that we have the joint density with observation $y$, as
\begin{eqnarray*}
f(y,u) = \sum_{j=1}^\infty {\bf 1}\{u < \xi_j\} (w_j/\xi_j)\,f(y|\mu_j,c, \kappa)
\end{eqnarray*}
for some decreasing deterministic sequence $(\xi_j)$. See \cite{Kalli11} for details and choices of $(\xi_j)$. In particular, we use $\xi_j = \exp(-\gamma j)$, with $\gamma = 0.01$, but other sequences can be used.

The key here is that given $u$, the number of components is finite, and correspond to the indices $A_u = \{j: \xi_j > u \}$, such that
\begin{eqnarray*}
f(y|u) = \sum_{j \in A_u} (w_j/\xi_j)\,f(y|\mu_j,c, \kappa).
\end{eqnarray*}
Next we introduce the component indicator $d$ for each observation, which tells us from which component available the observation came from, resulting in the joint density with $(y,u)$ as
$$f(y,u,d)={\bf 1}(u<\xi_d)\,(w_d/\xi_d)\,f(y|\mu_d,c,\kappa).$$ 
The joint likelihood function, including latent variables, is then given by
\begin{eqnarray*}
\prod_{i=1}^n {\bf 1}\{u_i < \xi_{d_i}\} (w_{d_i}/\xi_{d_i}) \,f(y_i|\mu_{d_i},c, \kappa), 
\end{eqnarray*}
and it will be convenient to define $D = \max\{d_i\}$.

\subsubsection{The MCMC algorithm}

It turns out that the most complicated parameter to deal with is $\kappa$; surprisingly, since often the location parameter is one of the easiest to sample in a MCMC routine. Thus, we will initially describe the algorithm assuming $\kappa$ is known. The unknown quantities that need to be sampled from, under that scenario, are: $(\mu_j, v_j), j=1,2,\ldots$; $(d_i, u_i), i=1,\ldots,n$; $c$ and $M$. The sampling of these parameters and latent variables is as follows, obviously conditional on other parameters and variables being known in each case: 

\begin{enumerate}

\item Sample $u_i$ from the uniform distribution on $(0,\xi_{d_i})$, for $i=1,\ldots,n$. A useful summary to record here is $N=\max\{\xi^{-1}(d_i)\}$.

\item Sample from $M$ following the approach in \cite{escobarwest95}. Now $M$ only depends on the other model parameters through the number of distinct $(d_i)$ and the sampling goes as follows,
\begin{itemize}
\item sample $\nu \sim B(\widetilde{M}, n)$, where $\widetilde{M}$ is the current value for $M$.
\item sample $M$ from $ga(\alpha_M + k, \beta_M - \log\nu), $ 
\end{itemize}
where $k$ is the number of distinct $(d_i)$.

\item Sample $(\mu_j)$ for $ j=1, \ldots, N$ via Metropolis-Hastings (M-H). If there is at least one $d_i = j$ then we use the M-H step, otherwise the $\mu_j$ comes from the prior. The proposal $\mu_{j}^*$ is taken to be normal, centered at the current value, i.e.
$$
\mu_{j}^* = \mu_{j} + N(0,h_\mu). 
$$
The M-H acceptance criterion is given by:
\begin{eqnarray*}
q=\frac{ \prod_{\{d_i = j\}} f(y_i|\mu_{j}^*,c,\kappa) \pi(\mu_{j}^*) } { \prod_{\{d_i = j\}} f(y_i|\mu_j,c,\kappa) \pi(\mu_j) }, 
\end{eqnarray*}
where $\pi$ represents the prior distribution.
We accept the proposal with probability $\min\{1,q\}$.

\item Sample $c$ via Metropolis-Hastings. The proposal $c^*$ is obtained from: 
\begin{eqnarray*}
\log(c^*) = \log(c) + N(0,h_c). 
\end{eqnarray*}
The M-H acceptance criterion is given by:
\begin{eqnarray*}
q=\frac{ \prod_{i=1}^n f(y_i|\mu_{d_i},c^*,\kappa) \pi(c^*)Q(c^*;c) }{ \prod_{i=1}^n f(y_i|\mu_{d_i},c,\kappa) \pi(c)Q(c;c^*) }, 
\end{eqnarray*}
where $Q(c^*;c)$ denote a source density for a candidate draw $c^*$ given the current value $c$ in the sampled sequence. In this case, $Q(c^*;c) = 1/c^*$, and we accept the proposal with probability $\min\{1,q\}$.

\item Sampling from $d_i$ can be done directly since we have the probability mass function,
\begin{eqnarray*}
P(d_i = j) = \frac{w_j}{\xi_j}\,f(y_i|\mu_j,c,\kappa) \quad \mbox{for}\quad j = 1, \ldots, N_i,
\end{eqnarray*}
where  $N_i=\max\{j:u_i<\xi_j\}$.

\end{enumerate}

When the mode $\kappa$ is unknown, simply adding an extra step in the algorithm above is neither efficient nor viable. The reason for this is that $\kappa$ and the $(\mu_j)$ are highly correlated. Data values coming from cluster $j$ with $\mu_j > 0$ are typically larger then $\kappa$, while if $\mu_j < 0$, they are typically smaller. The consequence is that for fixed values of $\mu_j$, any proposals for $\kappa$ are prone to be rejected. 

\begin{figure}[h!]
	\centering
\includegraphics[width=8cm,height=3cm]{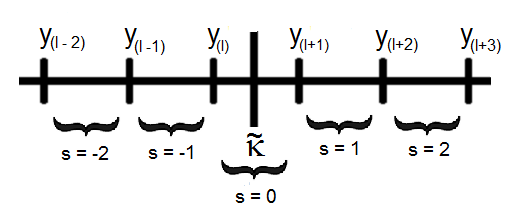} 
\caption{Toy example of initial value for $\kappa$ and possible values of $s$ in the MCMC algorithm.}
\label{fig:toy}
\end{figure}

The explanation for this is the following. Let $\{y_{(1)}, y_{(2)}, \ldots, y_{(n)}\}$ be the ordered data, and suppose the initial/current value for $\kappa$, written as $\kappa^*$, in the MCMC algorithm, places it between $y_{(l)}$ and $y_{(l+1)}$, for some $l = 1, \ldots, n-1$, as shown in Figure \ref{fig:toy}. It is likely that $\{y_{(1)},\ldots,y_{(l)}\}$ belong to clusters with negative $\mu_j$, and $\{y_{(l+1)},\ldots,y_{(n)}\}$ belong to clusters with positive $\mu_j$. This is true particularly for data which is close to $\kappa$, as from (\ref{withx}) we can see that they correspond to large values of $|\mu_j|$. These large values typically generate positive values of $x/z$ when $\mu_j > 0$, and negative values of $x/z$ when $\mu_j < 0$. 

Suppose we start the algorithm assigning to the data clusters whose $\mu_{d_i}$ have a sign which is negative for $y_i$ smaller than $\kappa$, and positive otherwise. If we now propose moving the value of $\kappa$ to, say, somewhere between $y_{(l+1)}$ and $y_{(l+2)}$, it means that $y_{(l+1)}$ will be incompatible with its corresponding cluster. The same happens if we move $\kappa$ in the opposite direction. More drastic moves make the problem more acute.

A solution for this would be to test moves of $\kappa$ and  $(\mu_{d_i})$ jointly in a Metropolis-Hastings step. That, however, would be very inefficient due to the possible high numbers of  $(\mu_j)$ involved. Our idea then is to test a new proposal for $\kappa$ jointly with moving the affected $y_i$, so the M-H step utilized to sample $\kappa$, and possibly some of the  $d_i$, is now described. 

To make the proposal, first sample auxiliary variable $s$ from a discrete uniform distribution in the interval $[-m,m]$, for a chosen $m \in {\rm I\!N}^+$. This sampled value will give the size and direction of the move we wish to propose for $\kappa$. 

As a toy example, suppose we choose $m=2$. Therefore, $s$ will be either $-2, -1, 0, 1$ or $2$, with the same probability. If $\tilde{\kappa}$ is placed between order statistics $y_{(l)}$ and $y_{(l+1)}$ we have that: if $s = -2$, $\kappa^*$ will be sampled uniformly within the interval $[y_{(l-2)},y_{(l-1)}]$; if $s = -1$, $\kappa^*$ will be sampled uniformly within the interval $[y_{(l-1)},y_{(l)}]$, and so forth, as illustrated in Figure $\ref{fig:toy}$. 

Note, however, that near the edges ($y_{(1)}$ and $y_{(n)}$) not every one of these movements can be made. For instance, in the example above, if $l = 2$, the movement proposed by $s = -2$ is not possible, since there is no interval $[y_0,y_1]$. In that case, $s$ is sampled uniformly between the possible values, which in this case would be $s = -1, 0, 1$ and $2$.   

Generalizing the notation, $\kappa^*$ will be sampled within the interval $[\kappa_a, \kappa_b]$, where $\kappa_a = \max \left\{ y_{(h+s)},y_{(1)} \right\}$, and $\kappa_b = \min \left\{y_{(h+s+1)},y_{(n)} \right\}$, with $h$ representing the rank of the order statistics of the $y$ which is immediately ranked lower than $\tilde{\kappa}$. In our toy example, $h = l$.

If $s = 0$, only a new proposal for $\kappa$ will be made, $\kappa^* \sim U(\kappa_a, \kappa_b)$; otherwise we will also test moving the observations which fall between $\tilde{\kappa}$ and $\kappa^*$ to a different cluster. If $\kappa^* > \tilde{\kappa}$, we propose moving these observation to the same cluster of $y_{(h)}$. If $\kappa^* < \tilde{\kappa}$, we propose moving them to the same cluster of $y_{(h+1)}$. In our example, if $s = 2$, it means proposing that $d^{o}[h+1]$ and $d^{o}[h+2]$ are equal to $d^{o}[l]$, where $d^{o}[l]$ represents the cluster to which $y_{(l)}$ belongs to. In the general case, we have:

\begin{itemize}
\item if $s<0$, we propose that $d^{o}[h+s+1],\ldots,d^{o}[h]$ are equal to $d^{o}[h+1]$;
\item if $s>0$, we propose that $d^{o}[h+1],\ldots,d^{o}[h+s]$ are equal to $d^{o}[h]$. 
\end{itemize}

The Metropolis rate is given by:
\begin{eqnarray*}
 q = \frac{p(\kappa^*,d^*|\cdot)}{p(\kappa,d|\cdot)} \propto \frac{\prod_{i=1}^n f(y_{(i)}|\mu_{d_{pi}},c,\kappa^*) \pi(\kappa^*)Q(\kappa^*,d^*;\kappa,d)}{ \prod_{i=1}^n f(y_{(i)}|\mu_{d_i},c,\kappa) \pi(\kappa)Q(\kappa,d;\kappa^*,d^*)},
\end{eqnarray*}
where 
\begin{eqnarray*}
Q(\kappa^*,d^*;\kappa,d) = \frac{(\min\{n,h+m\} - \max\{0,h-m\})^{-1}}{\kappa_b - \kappa_a},
\end{eqnarray*}
and, analogously, 
\begin{eqnarray*}
Q(\kappa,d;\kappa^*,d^*) = \frac{(\min\{n,h+s+m\} - \max\{0,h+s-m\})^{-1}}{y_{(h+1)} - y_{(h)}}.
\end{eqnarray*}
We accept the proposal with probability $\min\{1,q\}$.

\subsection{Examples with simulated data}

To test our model and the proposed MCMC algorithm, we simulate two artificial data-sets, and predict their densities. The idea here is verify if we can efficiently reconstruct the densities, which in this case are known. The simulated models are given bellow:

\begin{itemize}

\item[A.] $y$ is sampled from $N(100,100^2)$;

\item[B.] $y$ is sampled from $ga (3,10)$.

\end{itemize}

\begin{figure}[h!]
	\centering
		\includegraphics[width=8cm,height=4cm]{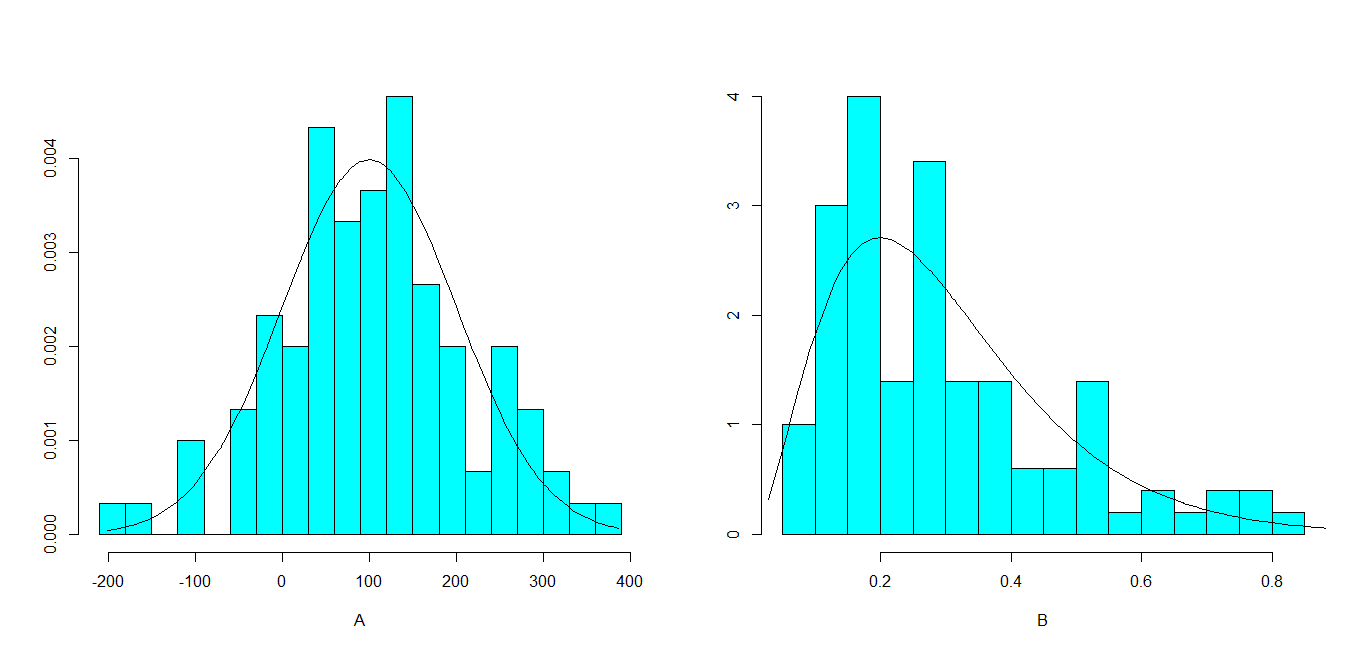} 
			\caption{Histograms based on $n=100$ points sampled from models A and B. The black line represents their respective densities.}
				\label{simuldata}
\end{figure}

Data-sets of size $n=100$ were simulated from models A and B. Figure \ref{simuldata} shows the histogram of these data-sets and the density of the distribution they were generated from. All figures included in this article were produced via the software R (\cite{R14}). 

The two sets of observations are modeled through \ref{inftysum}, with the following specifications for the parameters of the prior distributions: $\sigma_\mu^2 = 10$, $\sigma_\kappa^2=10000$, $\alpha_c = \beta_c = 0.1$, and $\alpha_M = \beta_M = 0.01.$  

For each data-set, the algorithm proposed in section $4$ was applied to sample from the posterior distribution of the model parameters. The MCMC was implemented in the software Ox version 7.0 (\cite{doornik08}), with $T=300000$ iterations and a burn in of $10000$. Due to auto-correlation in the MCMC chains, the samples were recorded at each $100$ iterations, leaving us with samples of size $2900$ to perform inference. 

\begin{figure}[h!]
	\centering
		\includegraphics[width=8cm,height=4cm]{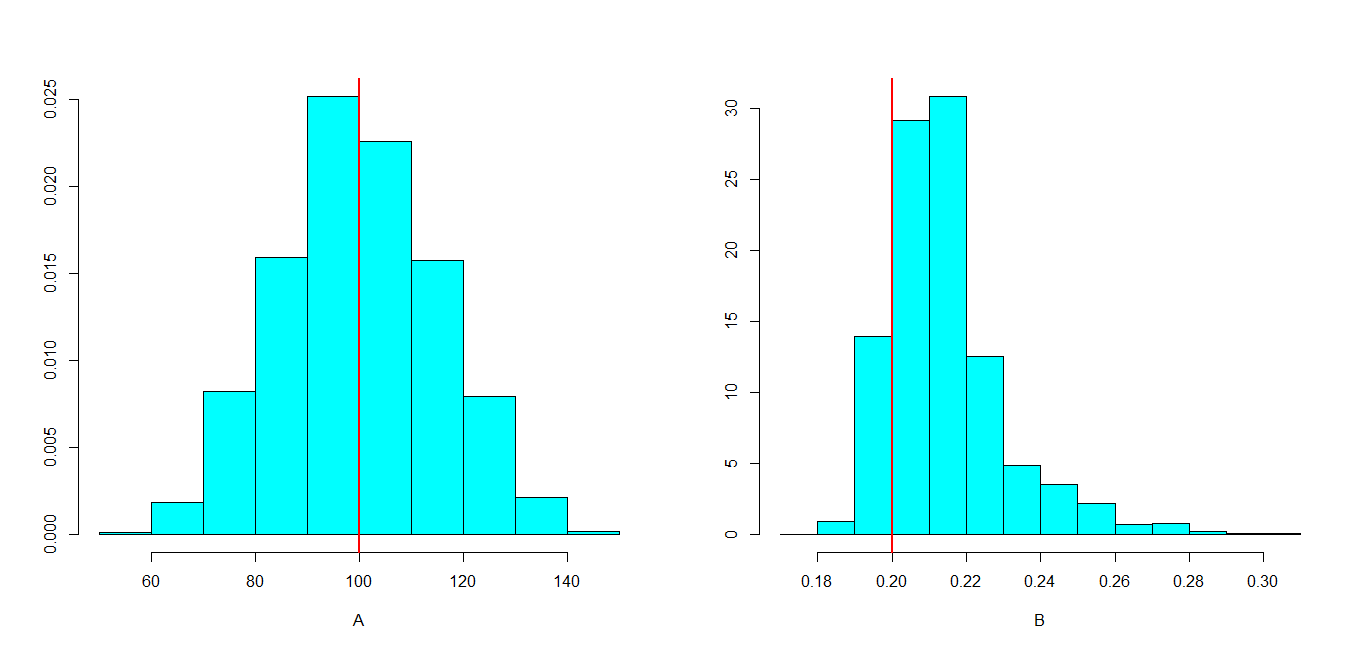} 
			\caption{Histograms based on $2900$ samples from the posterior of parameter $\kappa$ under observations from models $A$ and $B$. Red lines indicate the true values of the parameter.}
				\label{histA}
\end{figure}

Figure \ref{histA} shows the histogram of the sample from the posterior of parameter $\kappa$ under data-sets A and B. Note that in both cases the real value of the parameter is well estimated, falling close to the mode of the estimated posterior density.

\begin{figure}[h!]
	\centering
		\includegraphics[width=8cm,height=4cm]{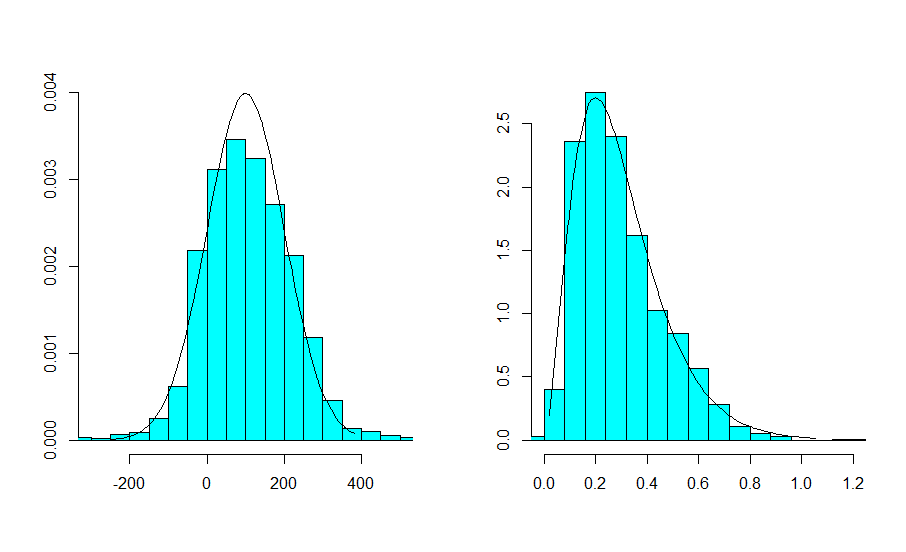} 
\caption{Histograms based on $2900$ samples from the predictive distribution under observations from models $A$ and $B$. Black line indicates the true density.}
\label{predA}
\end{figure}

Figure \ref{predA} shows the histograms of the predictive distributions under both data-sets compared with their real density. It can be seen that in both cases the predictive distribution is close to the real density, showing the flexibility of the model and validating the methodology. 

\section{Models for multivariate densities}

This section is divided in two sub-sections. Firstly, in sub-section \ref{multuni}, we discuss different definitions of multivariate unimodality that can be found in the literature. In sub-section \ref{multext} we propose a class of multivariate unimodal densities, extending the univariate densities of section \ref{unimoddens}.

\subsection{Multivariate unimodality} \label{multuni}

The concept of unimodality is well established for univariate distributions, and, following Khintchine's representation (\cite{Khint38}), it can be defined in different but equivalent ways. It is not straightforward, however, to extend the notion of unimodality to higher dimensional distributions, as different definitions of unimodality are not equivalent when extended. 

The first attempts to define multivariate unimodality were made for symmetric distributions only, by \cite{anderson55} and \cite{sherman55}. Again under symmetry, \cite{dharma76} introduced the notion of {\it convex unimodality} and {\it monotone unimodality}. The authors also defined {\it linear unimodality}, which can be applied to asymmetric distributions. A random vector $(Y_1, \ldots, Y_n)$ is {\it linear unimodal} about $0$ if every linear combination $\sum_{i=1}^ n{a_i Y_i}$ has a univariate unimodal distribution about $0$. This, however, is not a desired definition, as pointed out by \cite{dharma76} themselves, as the density of such may not be a maximum at the mode of univariate unimodality.  

Further, \cite{olshen70} define {\it $\alpha-$unimodality} about $0$ for a random variable $Y$ in ${\rm I\!R}^n$ when for all real, bounded, nonnegative Borel function $g$ on ${\rm I\!R}^n$ the function $t^{\alpha}E(g(tY))$ is non-decreasing as $t$ increases in $[0,\infty)$, where $E$ denotes the expectation with respect to the density of $Y$. If $X$ has a density function $f$ with respect to the Lebesgue measure $\mu_n$ on ${\rm I\!R}^n$, then $X$ is {\it $\alpha-$unimodal} if and only if $t^{n-\alpha}f(ty)$ is decreasing in $t \geq 0$. The distribution of a random vector $(Y_1, \ldots, Y_n)$ is said to be {\it star unimodal} about $0$, according to \cite{dharma76}, if it is $n$-unimodal in the sense of \cite{olshen70}. 

According to \cite{dharma76}, a {\it linear unimodal} distribution need not be {\it star unimodal}, and vice versa. Thus there is no implied relationship between {\it star} and {\it linear unimodality}. 

Another useful definition of unimodality is given by \cite{devroye97}: A multivariate density $f$ on ${\rm I\!R}^d$ is {\it orthounimodal} with mode at $m = (m_1,\ldots,m_d)$ if for each $j$, $f(y_1,\ldots,y_d)$ is a decreasing function of $y_j$ as $y_j \rightarrow \infty$ for $y_j \geq m_j$, and as $x_j \rightarrow -\infty$ for $y_j \leq m_j$, when all other components are held fixed. 

{\it Orthounimodal} densities are  widely applicable. They form a robust class in the sense that all lower-dimensional marginals of
{\it orthounimodal} densities are also {\it orthounimodal}. Also, they are {\it star unimodal}, according to \cite{devroye97}, and most bivariate densities are either {\it orthounimodal}, or {\it orthounimodal} after a linear transformation. Narrower notions than {\it orthounimodality} can also be explored, and they are discussed in \cite{dharma88}. 

For the two dimensional case only, \cite{shepp62} generalizes Khintchine's stating that the distribution function of $(Y_1, Y_2)$ is unimodal if and only if it can be written as
\begin{equation}
 (Y_1, Y_2) = (X_1 Z_1, X_2 Z_2), \label{bikhin}
\end{equation}
where $(X_1, X_2)$ and $(Z_1, Z_2)$ are independent and $(X_1, X_2)$ is uniformly distributed in $[0,1]$. For higher dimensions, Khintchine's representation was generalized by \cite{kanter77} for the symmetric case. He defines symmetric multivariate unimodal distributions on ${\rm I\!R}^d$ as generalized mixtures (in the sense of integrating on a probability measure space) of uniform distributions on symmetric, compact and convex sets in ${\rm I\!R}^d$. In this paper we will use a form of (\ref{bikhin}) while guaranteeing orthounimodality.

Comprehensive reviews on multivariate unimodality can be found in \cite{dai82} and \cite{dharma76}. Also \cite{Kouvaras08} reviews the literature and introduces and discusses a new class of nonparametric prior distributions for multivariate multimodal distributions. 

\subsection{Proposed model} \label{multext}
To \hspace{1 pt} extend \hspace{1 pt} the \hspace{1 pt} univariate \hspace{1 pt} unimodality \hspace{1 pt} proposed \hspace{1 pt} in \hspace{1 pt} section \hspace{1 pt} \ref{unimoddens},\hspace{1 pt}  we \hspace{1 pt} start \hspace{1 pt} by de- \\fining \hspace{1 pt} marginal \hspace{1 pt} variables \hspace{1 pt} $Y_1, \hspace{1 mm} Y_2, \hspace{1 pt} \ldots, \hspace{1 pt} Y_d$ \hspace{1 pt} via \hspace{1 pt} $Y_l \hspace{1 pt} = \hspace{1 pt} X_l/Z_l$, \hspace{1 pt} where \hspace{1 pt} $X_l \hspace{1 pt} \sim \hspace{1 pt} U(0,1)$ and $Z_l \sim N(\mu_l,\mu_l^2/c_l)$, for $l=1,2,\ldots,d$. The dependence between variables $Y = (Y_1, Y_2, \ldots, Y_d)$ can be obtained imposing dependence to either $Z = (Z_1,\ldots,Z_d)$ or $X = (X_1,\ldots,X_d)$, or both.

Our first attempt was to allow dependence only in $Z$ through a multivariate normal distribution with a general variance-covariance matrix. This way our construction would be following the generalization of Khintchine's representation made by \cite{shepp62} for the bivariate case. The dependence obtained between $Y$ under that construction, however, for a bivariate setting, is limited. Better results were obtained when imposing dependence in $X$ instead, through a Gaussian copula. 

\begin{figure}[h!]
	\centering
		\includegraphics[width=12cm,height=7cm]{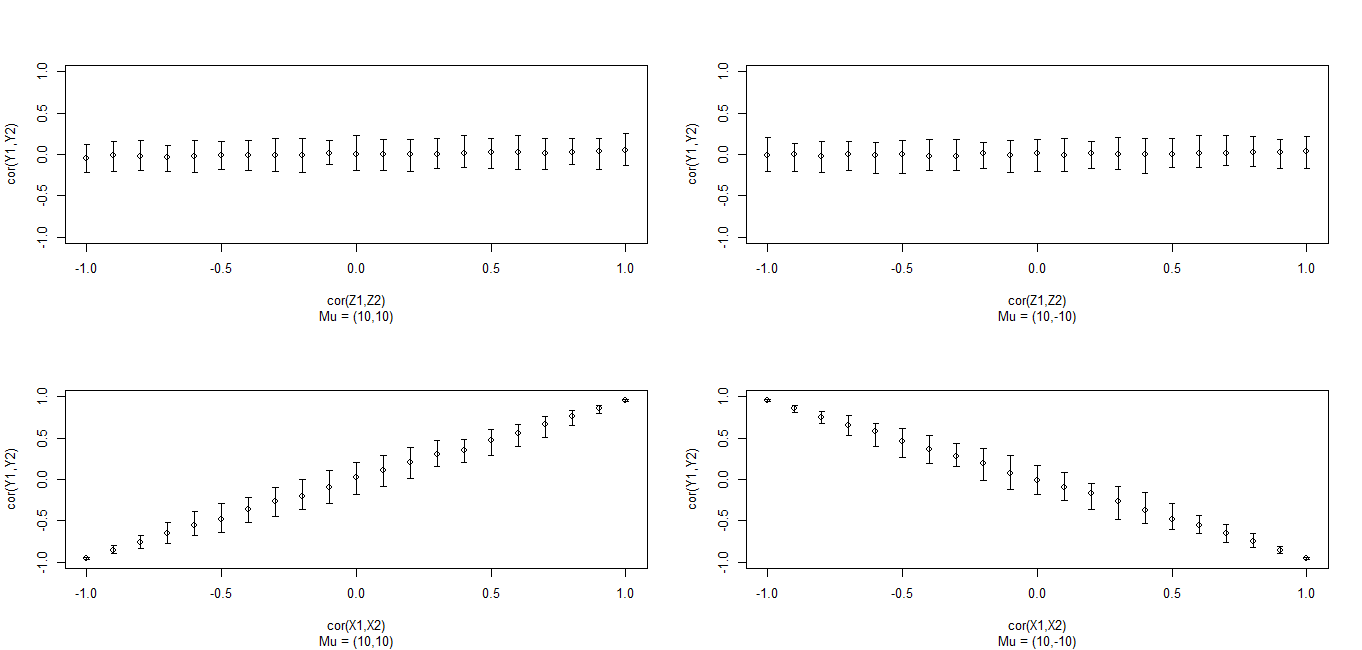} 
		 \caption{95\% confidence intervals of the correlation obtained between $(Y_1,Y_2)$ when varying the correlation between $(Z_1,Z_2)$ (top) and $(X_1,X_2)$ (bottom) respectively, through the interval $\{-1,-0.9,-0.8,\ldots,0.9,1\}$.}
				\label{correlations}
\end{figure}

To demonstrate: Figure \ref{correlations} shows the approximate 95\% confidence intervals of the correlations between $Y$ when varying the correlations between $Z$ (top) and $X$ (bottom) respectively, through the interval $\{-1,-0.9,-0.8,\ldots,0.9,1\}$. These confidence intervals were found by the simulation of $100$ samples of size $100$. The figure was constructed considering $\mu=(10,10)$ and $\mu=(-10,10)$. Similar results to the ones with $\mu = (10,10)$ were found for $\mu = (-10,-10)$. In the same way, $\mu=(10,-10)$ and $\mu=(-10,10)$ present similar results. The magnitude of $\mu$ did not seem to change this output significantly. It can be seen that when imposing dependence through $Z$, the correlation between $Y$ do not vary much beyond the interval $[-0.2,0.2]$, while the correlation between the $(Y_1,\ldots,Y_d)$ and the correlation between the $(X_1,\ldots,X_d)$ was found to be similar. 

Due to the results just described we opted to work with the second construction: i.e. $Y_l = X_l/Z_l$, $Z_l \stackrel{ind}\sim N(\mu_{lj},\mu_{lj}^2/c_l), l = 1,\ldots, d$, and $X = (X_1, X_2, \ldots, X_d)$ are modeled through the Gaussian copula with correlation matrix $\Sigma$. Once again, the mode at $\kappa = (0,\ldots,0)$ can be exchanged, substituting $Y$ for $(Y - \kappa)$ where $\kappa$, as well as $Y$, is now a $d-$dimensional vector. 

Given $x$, $\mu_j = (\mu_{1j},\ldots, \mu_{dj})$, $c = (c_1,\ldots, c_d)$ and $\kappa = (\kappa_1,\ldots, \kappa_d)$, the observations $y = (y_1, y_2, \ldots, y_d)$ are independent, and $f(y|x,\mu_{j},c,\kappa)$ is given by
$$
f(y|x,\mu_{j},c,\kappa) = \prod_{l=1}^d (x_l/y_l^2)g_l(x_l/y_l|\mu_{lj},c_l,\kappa_l).
$$
The marginal model, $f(y|\mu_{j},c,\kappa)$, can be obtained through the integral
\begin{equation}
f(y|\mu_{j},c,\kappa) = \int_{[0,1]^d} \prod_{l=1}^d (x_l/y_l^2)g_l(x_l/y_l|\mu_{lj},c_l,\kappa_l) \d c(x_1, \ldots, x_d|\Sigma), \label{intmult}
\end{equation}
where $c(x_1, \ldots, x_d|\Sigma)$ represents the multivariate Gaussian copula with correlation matrix $\Sigma$:
\begin{equation*}
c(x_1, \ldots, x_d|\Sigma) = \frac{1}{\sqrt{|\Sigma|}} \exp \left\{ -\frac{1}{2} \left( \begin{array}{c} \Phi^{-1}(x_1) \\ \vdots \\ \Phi^{-1}(x_d) \end{array} \right) (\Sigma^{-1} - I) \left( \begin{array}{c} \Phi^{-1}(x_1) \\ \vdots \\ \Phi^{-1}(x_d) \end{array} \right) \right\}.
\end{equation*}

\begin{figure}[h!]
	\centering
		\includegraphics[width=11cm,height=7cm]{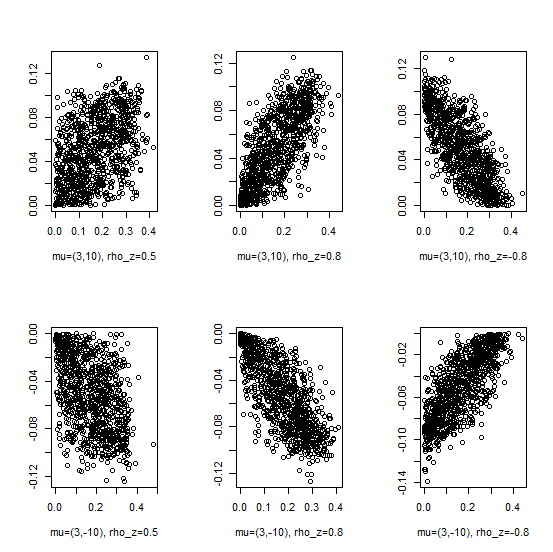} 
		 \caption{Dispersion Plot of samples $y_1$ vs $y_2$ of size $1000$ from the simulated bivariate example.}
				\label{example_multi_1}
\end{figure}

Note, however, that the dependence created between variables $Y$ is defined not only by the correlation matrix $\Sigma$, but also by the sign of the components of $\mu$. As an illustration, we sampled four bivariate data-sets, and examined the dispersion plots between $y_1$ and $y_2$. Figure \ref{example_multi_1} shows the dispersion plot between data sampled with $\mu_1 = 3$, and either $\mu_2 = 10$ or $\mu_2 = -10$, and with $\rho = \Sigma[1,2] = 0.5$, $\rho = 0.8$ or $\rho = -0.8$. It can be seen that positive dependence is created when $(\mu_1 \times \mu_2 \times \rho)$ is positive, and negative dependence is created otherwise. To help the model identification, we assume that $\rho \in [0,1]$. This restriction does not take away the model flexibility, and negative correlations between the variables can be captured by opposite signs in $\mu$.  

Let us now define the mixture of these densities, which is the unimodal multivariate model of interest. For observation $Y = (Y_1, \ldots, Y_d)$ the model is:  
\begin{equation*}
f(y) = \sum_{j=1}^\infty{w_j f(y|\mu_{j},c,\kappa)},
\end{equation*} 
where $f(y|\mu_{j},c,\kappa)$ is in ($\ref{intmult}$). Note that, by construction, the proposed distribution is multivariate {\it orthounimodal}, which is, therefore, also {\it star unimodal}. Unlike the univariate case, however, we cannot solve the integral in (\ref{intmult}) and obtain $f(y|\mu_{j},c,\kappa)$ analytically. Therefore, we must come up with a different algorithm to solve the sampling of $\kappa$. As a ``bridge'' to the multivariate case, this algorithm is first developed for the univariate case, and then extended to higher dimensions. The univariate ``bridge'' algorithm is presented in subsection \ref{bridge}. 

\subsection{MCMC algorithm for univariate ``bridge''} \label{bridge}

To mimic the situation in which the latent variable $x$ cannot be integrated out, we present in this section an algorithm which samples from $x = (x_1, \ldots, x_n)$, and samples from the other parameters given $x$. We consider the same prior distributions specified in section \ref{priors}. For $x$ we assume a uniform prior at $[0,1]^n$. The algorithm goes as follows:

\begin{enumerate}

\item Sample $u_i$ and obtain $N=\max\{\xi^{-1}(d_i)\}$ as previously in section \ref{priors};

\item Sample $M$ as previously in section \ref{priors};

\item Sample $(\mu_j)$ for $j = 1, \ldots, N$ as before, but using $f(y_i|x_i,\mu_j,c,\kappa)$ (proportional to \ref{withx}) instead of $f(y_i|\mu_j,c,\kappa)$ in the Metropolis ratio. This way, the M-H acceptance criterion is given by:
\begin{eqnarray*}
q=\frac{ \prod_{i=1}^n f(y_i|x_i,\mu_{d_i},c^*,\kappa) \pi(c^*) }{ \prod_{i=1}^n f(y_i|x_i,\mu_{d_i},c,\kappa) \pi(c) }. 
\end{eqnarray*}

\item The full conditional distribution of $c$ in this case has a closed form, and it is updated via Gibbs sampling. Given $y$, $x$, $\mu$ and $\kappa$, we have:
$$
c|x,\mu,\kappa,y \sim ga\left( \frac{n}{2} + \alpha_c, \frac{1}{2} \sum_{i=1}^n \left( \frac{x_i}{y_i - \kappa} - \mu_{d_i} \right)^2 + \beta_c \right).
$$  

\item Sample $(x,d,\kappa|\mu, c, y)$ in two stages: first sample from $f(d, \kappa|\mu, c, y)$ and then from $f(x|d, \kappa, \mu, c, y)$. $d$ and $\kappa$ are sampled as before, and $(x_i)$ for $ i=1, \ldots, n$, is sampled via rejection sampling, as follows:

\begin{itemize}

\item sample a proposal $\tilde{x_i} \sim U[0,1]$;
\item compute $f(\tilde{x}_i|c,\mu_{d_i},y_i,\kappa) \propto \tilde{x}_i \mbox{exp} \left\{ -\frac{c}{2 \mu_{d_i}^2} \left( \frac{\tilde{x}_i}{y_i - \kappa} - \mu_i \right) \right\}  $ 
\item compute the value $\hat{x}$ which maximizes the function: 
\begin{equation*}
 \hat{x} = \min \left\{ 1, \frac{ \mu_{d_i}(y_i - \kappa) }{2}+ \left| (y_i - \kappa) \sqrt{ \mu_{d_i}^2/c + 0.25 \mu_{d_i} } \right| \right\}. 
\end{equation*}
\item accept $\tilde{x_i}$ with probability $\min \left\{ 1, \frac{f(\tilde{x_i}|c,\mu_{d_i},y_i,\kappa)}{f(\hat{x_i}|c,\mu_{d_i},y_i,\kappa)} \right\} $, or go back to the first step.

\end{itemize}
\end{enumerate}

The results obtained under this algorithm were similar to those obtained from the previous one, despite being considerably slower. The new algorithm, however, can be easily extended to the multivariate case. In this paper we will discuss the algorithm and results obtained for the bivariate case, and leave higher dimensionality for future work.   

\subsection{MCMC algorithm for bivariate observations} \label{bivariate}

In this section we present the algorithm developed to handle bivariate observations ($p=2$). Under this scenario, the unknown quantities are: 
$(\mu_j = (\mu_{1j},\mu_{2j}), v_j), j=1,2,\ldots$; $(d_i, u_i), i=1,\ldots,n$; $c = (c_1, c_2)$; $M$; $\kappa = (\kappa_1, \kappa_2)$, $x_{li}, l=1,2, i=1,\ldots,n$, and also the correlation parameter in the bivariate Gaussian Copula ($\rho$). As pointed out in Section \ref{multext}, without compromising the model flexibility, we can assume that $\rho \in [0,1]$. Therefore we assigned a $U[0,1]$ prior for this parameter. 

The algorithm proposed for the bivariate case is given below:

\begin{enumerate}

\item Sample $u_i$ and obtain $N=\max\{\xi^{-1}(d_i)\}$ as previously;

\item Sample $M$ as previously;

\item Sample $(\mu_{1j})$ and $(\mu_{2j})$, independently for $j = 1, \ldots, N$, following the same idea as in (\ref{bridge}); 

\item Sample $c_1$ and $c_2$ via Gibbs sampling. Given $y$, $x$, $\mu$, and $\kappa$, we have:
$$
c_l|x,\mu,\kappa \sim ga \left( \frac{n}{2} + \alpha_c, \frac{1}{2} \sum_{i=1}^n \left( \frac{x_{li}}{y_{li} - \kappa} - \mu_{l,d_i} \right)^2 + \beta_c \right), l=1,2.
$$  

\item Sample $\kappa_1$ and $\kappa_2$ independently, using a similar algorithm as previously. Note that every time $\kappa_1$ and $\kappa_2$ are updated, some elements of $d$ might also be changed.
 
\item Sample $d_i$ directly from its probability mass function,
\begin{eqnarray*}
P(d_i = j) = \frac{w_j}{\xi_j}\,f(y_{1i}|\mu_{1j},c,\kappa_1)f(y_{2i}|\mu_{2j},c,\kappa_2) \mbox{ for }j = 1, \ldots, N_i,
\end{eqnarray*}
where  $N_i=\max\{j:u_i<\xi_j\}$.

\item Sample $(x_1, x_2)$ via rejection sampling. Here we consider two possible algorithms:

\begin{itemize}

\item[7.1.] 

\item Sample from the copula $C(x^*_{1i},x^*_{2i})$: 
$$
(\Phi^{-1}(x^*_1),\Phi^{-1}(x^*_2))^T \sim N \left( \left( \begin{array}{c} 0 \\ 0 \end{array} \right), 
\left( \begin{array}{cc} 1 & \rho \\  \rho & 1 \end{array} \right) \right); 
$$

\item compute 
$$f(x^*_{1i},x^*_{2i}|c,\mu_{d_{1i}},y_{1i},\mu_{d_{2i}},y_{2i},\kappa) \propto \prod_{j=1}^2{x^*_{ji} \mbox{exp} \left\{ -\frac{c}{2 \mu_{d_{ji}}^2} \left( \frac{\tilde{x}_{ji}}{y_{ji} - \kappa} - \mu_{ji} \right) \right\}}; $$ 

\item compute the values $\hat{x}_1$ and $\hat{x}_2$ which maximize the function above:
\begin{equation*}
 \hat{x}_j = \min \left\{ 1, \frac{ \mu_{d_{ji}}(y_{ji} - \kappa) }{2}+ \left| (y_{ji} - \kappa) \sqrt{ \mu_{d_{ji}}^2/c + 0.25 \mu_{d_{ji}} } \right| \right\}, j=1,2;
\end{equation*}

\item accept $(x^*_{1i}, x^*_{2i})$ with probability $$\min \left\{ 1, \frac{f(x^*_{1i},x^*_{2i}|c,\mu_{d_{1i}},y_{1i},\mu_{d_{2i}},y_{2i},\kappa)}{f(\hat{x}_{1i},\hat{x}_{2i}|c,\mu_{d_{1i}},y_{1i},\mu_{d_{2i}},y_{2i},\kappa)} \right\},$$ or go back to the first step.

\bigskip

\item[7.2.] 

\item Sample $x^*_{ji}$ from the function 
$$
f(x^*_{1i},x^*_{2i}|c,\mu_{d_{1i}},y_{1i},\mu_{d_{2i}},y_{2i},\kappa) \propto  {x^*_{ji} \mbox{exp} \left\{ -\frac{c}{2 \mu_{d_{ji}}^2} \left( \frac{x^*_{ji}}{y_{ji} - \kappa} - \mu_{ji} \right) \right\}},
$$
through adaptive rejection sampling (\cite{gilkswild91});  

\item compute $c(x^*_{1i},x^*_{2i})$;

\item compute the values $\hat{x}_1$ and $\hat{x}_2$ which maximize the copula: $\hat{x}_1 = \rho x_2$ and $\hat{x}_1 = \rho x_1$;

\item accept $(x^*_{1i}, x^*_{2i})$ with probability $$\min \left\{ 1, \frac{c(x^*_{1i},x^*_{2i})}{c(\hat{x}_{1i},\hat{x}_{2i})} \right\},$$ or go back to the first step.
\end{itemize}

The first algorithm proposed to sample from $x$ leads to faster convergence. However, it can be very slow at times, requiring a large amount of simulations until acceptance. We combined the two algorithms in the following way: Sample from 7.1 until it either accepts the proposal or reaches a set limited number of trials. If the latter occurs, sample from $x$ through algorithm 7.2.    

\item Sample $\rho$ via Metropolis-Hastings. The proposal $\rho^*$ is obtained from: 

$$
\log \left( \frac{\rho^*}{1 - \rho^*} \right) = \log \left( \frac{\rho}{1 - \rho} \right) + N(0,h_\rho^2), 
$$
for a suitable choice of $\sigma_\rho^2.$

The M-H acceptance criterion is given by:
\begin{eqnarray*}
q=\frac{ f(\rho^*|x_1,x_2) Q(\rho^*;\rho) }{ f(\rho|x_1,x_2) Q(\rho;\rho^*) }, 
\end{eqnarray*}
where $Q(\rho^*;\rho) = 1/(\rho^*(1- \rho^*))$. We accept the proposal with probability $\min\{1,q\}$.

\end{enumerate}

\subsection{Examples with simulated data}

In this section we present the results obtained for a simulated bivariate data-set of size $n=100$, from a bivariate Normal distribution: $y \sim N(\kappa, \Omega)$ with $\kappa = (30, 60)$, and $\Omega = 10 \left( \begin{array}{cc} 1 \quad \rho_y \\ \rho_y \quad 1 \end{array} \right)$, with $\rho_y = 0.5$. 

The algorithm proposed in section \ref{bivariate} was applied to sample from the posterior distribution of the model parameters and obtain samples of the predictive distributions. The algorithm was run for $T=75000$ iterations, with a burn in of $5000$. Due to auto-correlation in the MCMC chains, samples were recorded at each $50$ iterations, leaving a sample of size $1400$ to perform inference.

\begin{figure}[h!]
	\centering
 		 \includegraphics[width=10cm,height=5cm]{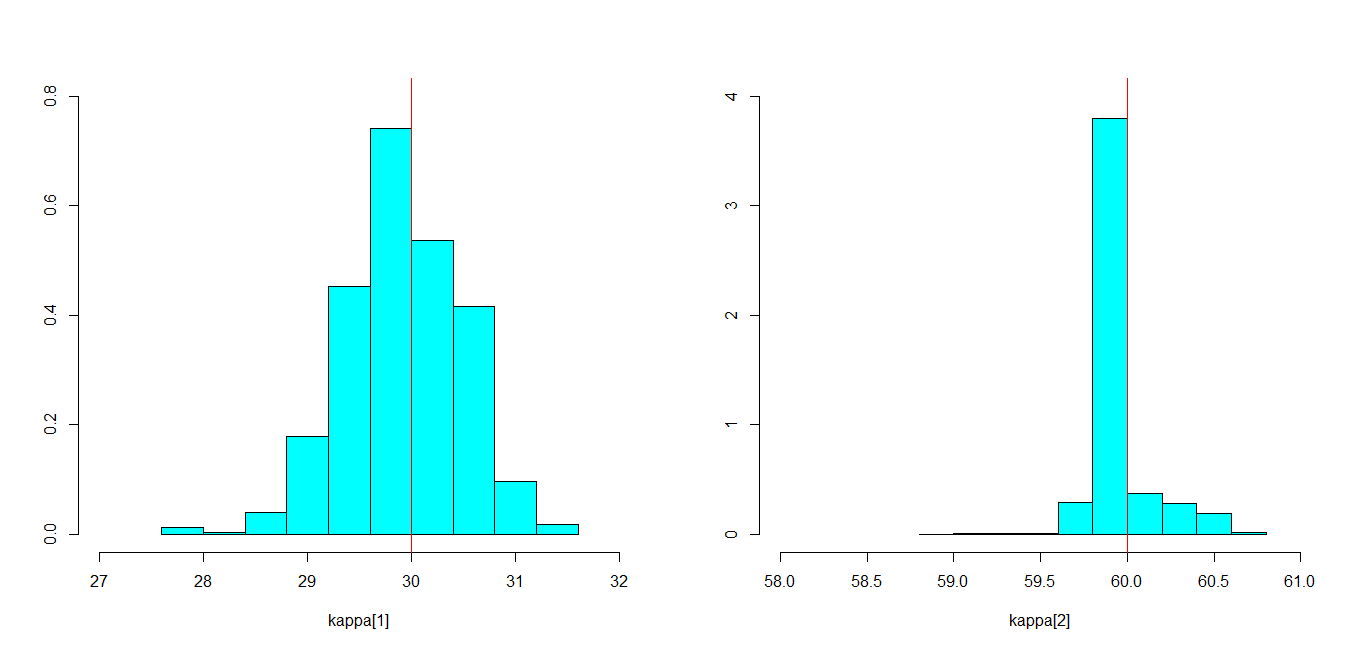}
		 \caption{Histograms of the posterior of $\kappa$ under simulated bivariate model. Red lines indicate true values.}
				\label{k_bi_2}
\end{figure}

\begin{figure}[h!]
	\centering
		\includegraphics[width=12cm,height=9cm]{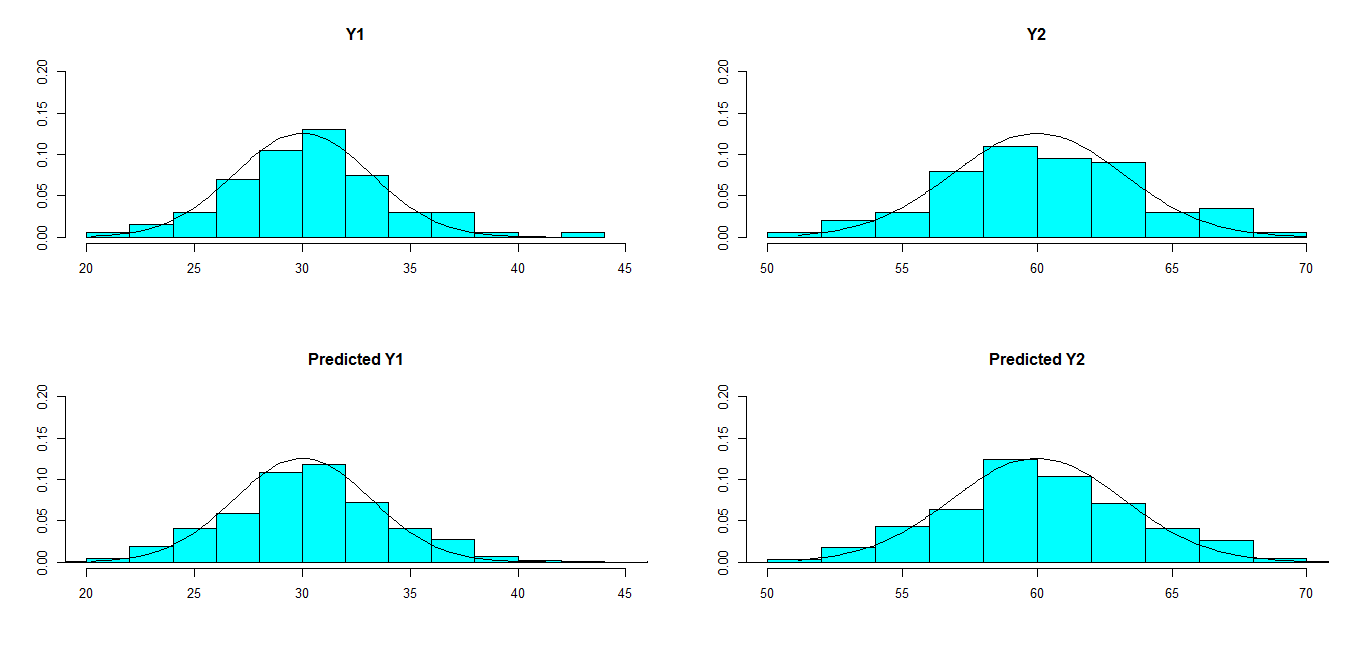} \\
		 \caption{Histograms comparing the original simulated samples (size 100) to the predicted samples under the bivariate model. Black lines represent the true density.}
				\label{hist_bi_2}
\end{figure}

Figure \ref{k_bi_2} shows the histograms of the posterior densities of the components of parameter $\kappa$. It can be seen that this parameter was reasonably well estimated, with the true value being close to the posterior mode. Figure \ref{hist_bi_2} presents a comparison between the histograms of the simulated samples of $y_1$ and $y_2$, and the ones predicted through the proposed MCMC algorithm, showing that the predictions seem close to what was expected.

\begin{figure}[h!]
	\centering
		\includegraphics[width=8cm,height=5cm]{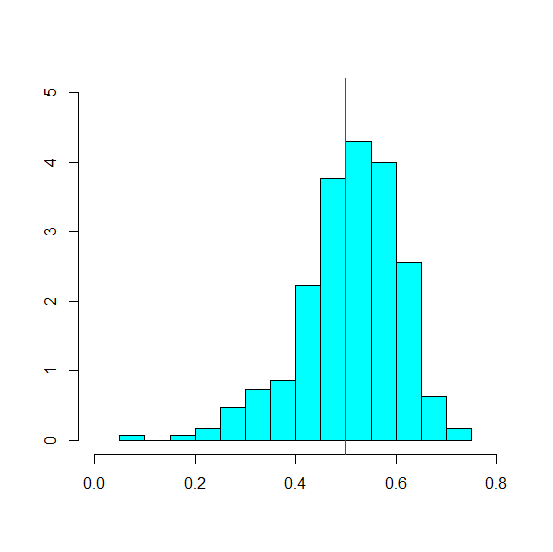} 
		 \caption{Histogram of the estimated correlation between samples of size $100$ of predicted $(y_1,y_2)$. The red line indicates the true correlation.}
				\label{cor_biB}
\end{figure}

We also wish to verify if the dependence between $y_1$ and $y_2$ was preserved in the predictions. To do this analysis, we computed the correlation at every $100$ samples of the predicted $(y_1,y_2)$. This time we used the full $T=70000$ sampled values, ending up with a sample of size $700$ correlations. Figure \ref{cor_biB} compares the histogram of these correlations, which can be seen as a proxy of the posterior distribution of $\rho_y$, to the real values (in red), showing good predictions. 

\section{Boston Housing Data}

\begin{figure}[h!]
\begin{center}
	\centering
		\includegraphics[width=11cm,height=4cm]{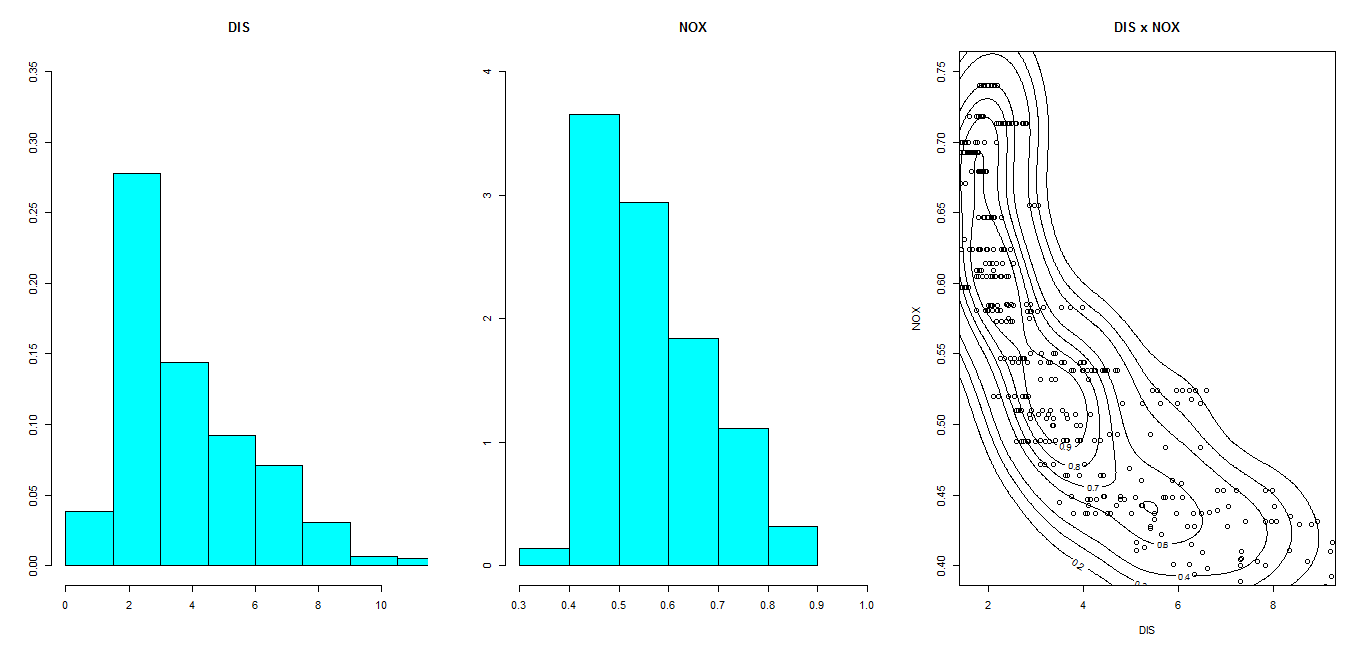} \\
		 \caption{Histograms of variables DIS and NOX, and contour plot of DIS vs NOX from the Boston Housing database.}
				\label{realdata}
\end{center}
\end{figure}

As an application, we work with a part of the Boston Housing data-set created by \cite{harrison78} and taken from \cite{Lichman13}. This database comprises $506$ cases of $12$ variables concerning housing values in the suburbs of Boston, and it has been used in many applications, especially in regression and machine learning, such as \cite{belsley80} and \cite{quinlan93}. We chose to work with two variables of the database: nitric oxides concentration (parts per 10 million) at the household (NOX) and weighted distances to five Boston employment centers (DIS). Exploratory analysis points towards the unimodality of the joint density of NOX and DIS, as can be seen by the contour plot displayed in Figure \ref{realdata}. This contour plot also shows a negative, unusually shaped, dependence between these variables. Histograms of observations NOX and DIS are also presented in Figure \ref{realdata}, showing that both variables have a non-normal unimodal shape. 

Our objective in this application is to illustrate the flexibility of our approach in capturing the form of this bivariate distribution and the oddly shaped dependence between NOX and DIS. We worked with a sub-set of $n=100$ observations and applied the proposed bivariate model with the same priors used for the toy examples. Again, the algorithm was run for $T=75000$ iterations, with a burn in of $5000$, and recorded at every $50$ iteration, totalizing a sample of size $1400$ for inference.

\begin{figure}[h!]
	\centering
		\includegraphics[width=11cm,height=4cm]{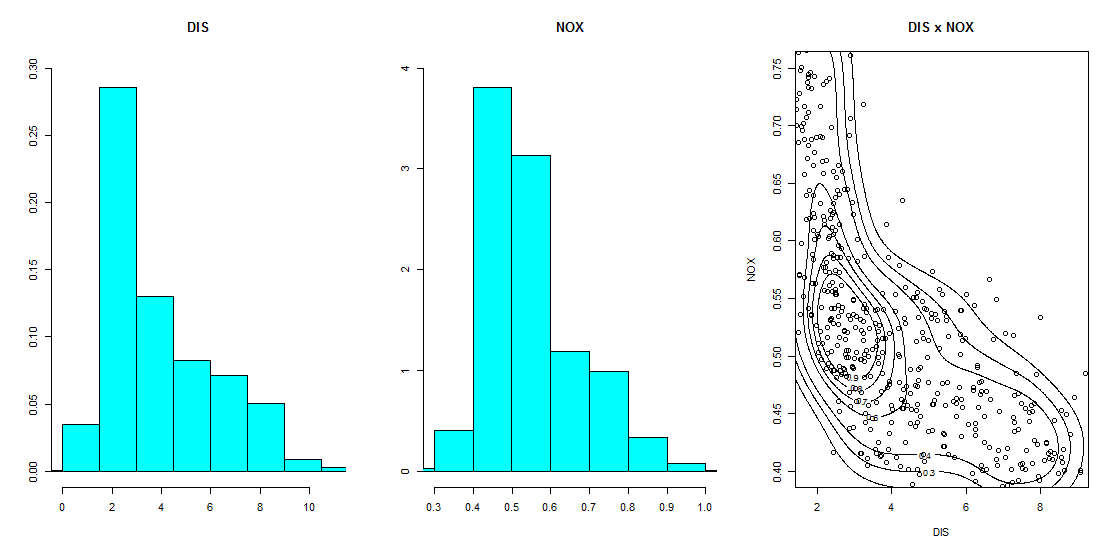} \\
		 \caption{Predicted histograms of DIS and NOX, and predicted contour plot of DIS vs NOX under the proposed bivariate mixture model.}
				\label{preddata}
\end{figure}

Figure \ref{preddata} shows the histograms of the predictive density of NOX and DIS, and the contour plot made with $500$ points of the predictive distribution. We observe a high similarity between the real and the predicted histograms and dispersion plots, and can conclude that the proposed methodology was able to capture well the behavior of the data for this example.

Note that our approach also allows for predictions of one variable given the other after samples of both had been previously observed.  As a second exercise, we analyze the predictive capacity of our model, compared to a more usual linear alternative. If the objective is to predict nitric oxides concentration based on weighted distances to employment centers, a regression model would probably be considered. As can be seen in Figure \ref{realdata}, however, the data needs transformation for the normal linear regression model assumptions to hold adequately. After exploratory analysis, we propose the following regression model:
\begin{eqnarray*}
\mbox{NOX}_i^{-1} &=& \alpha + \beta \mbox{log}(\mbox{DIS}_i) + e_i, \quad i = 1,\ldots,n, \\
e_i &\sim& N(0,\sigma_e^2),
\end{eqnarray*}
with vague Normal priors for $\alpha$ and $\beta$, and a vague inverse Wishart prior for $\sigma_e^2$.

\begin{figure}[h!]
		\includegraphics[width=11cm,height=4cm]{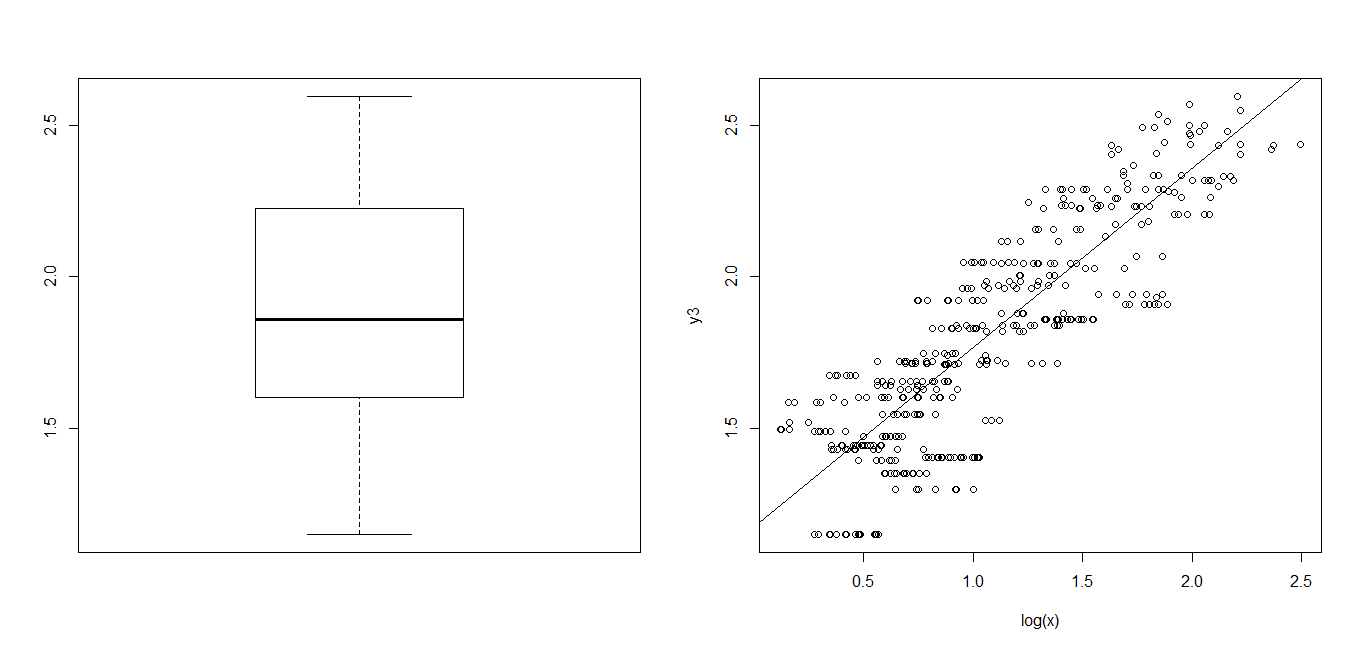} \\
		 \caption{Histogram of NOX$^{-1}$ and dispersion plot of log(DIS) vs NOX$^{-1}$ with regression line.}
				\label{tranfnorm}
\end{figure}

Figure \ref{tranfnorm} shows the box-plot of the response variable NOX$^{-1}$ and the dispersion plot between NOX$^{-1}$ and log(DIS), showing a linear dependence between both variables. The simple regression was fit using software OpenBugs 3.2.3 (\cite{thomas06}). 

\begin{figure}[h!]
	\centering
		\includegraphics[width=10cm,height=4cm]{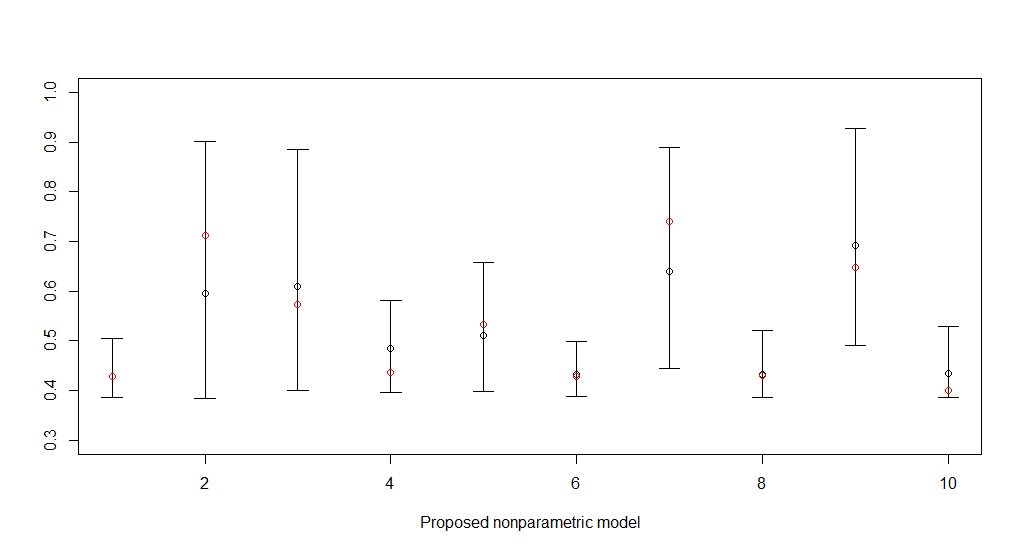} \\
		\includegraphics[width=10cm,height=4cm]{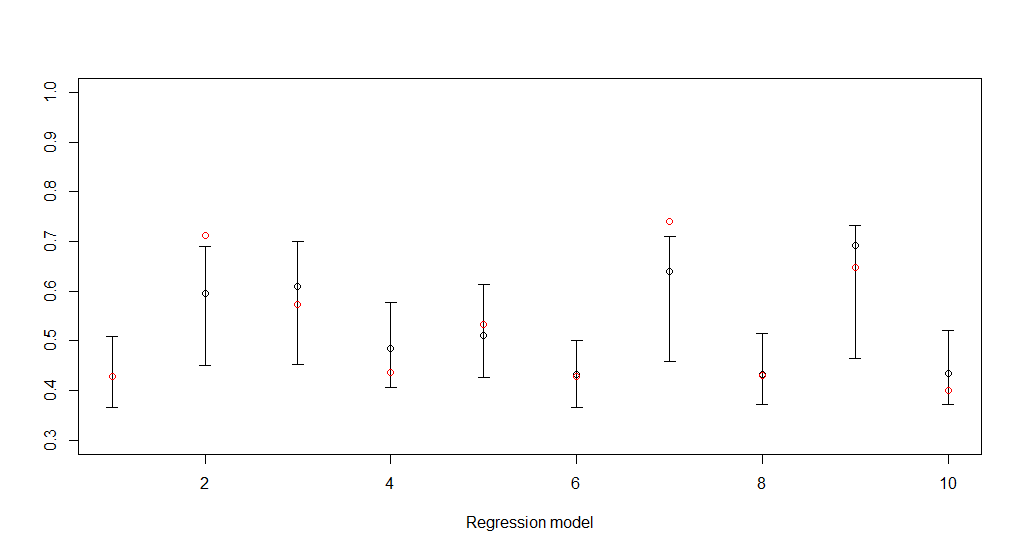} \\
		 \caption{95\% credible intervals and posterior medians of 10 ``future'' values of NOX being compared to their observed values (red dots) under the proposed nonparametric model (top) and a regression model (bottom).}				\label{preddata1}
\end{figure}

\begin{figure}[h!]
	\centering
		\includegraphics[width=10cm,height=5cm]{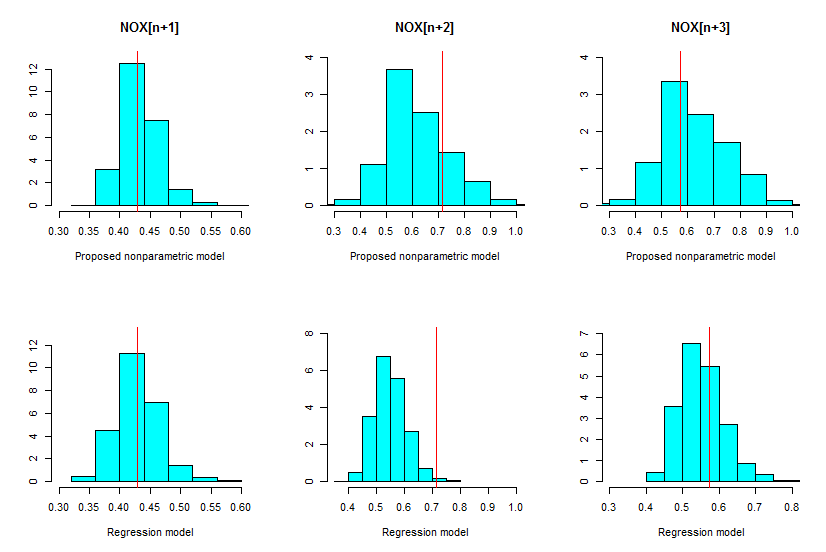} \\
		 \caption{histograms of the posterior of 3 ``future'' values of NOX under the proposed nonparametric model (top) and a regression model (bottom). Real observed values are indicated by red line.}				\label{preddata2}
\end{figure}

A comparison is then performed between predictions of $10$ values of NOX given DIS, after observing $n=100$ cases of both DIS and NOX, under both models. Figure \ref{preddata1} presents the 95\% credible intervals and posterior medians under the proposed nonparametric model and the simple regression model, being compared with the real values. Figure \ref{preddata2} presents the histograms of the posterior distribution of the first three predicted values. As is typical, nonparametric methods produce larger credible intervals than the parametric counterparts, yet are typically centered in the true values. The parametric versions however can simply be wrong. 

\section{Final remarks}

In this paper we have proposed a new class of unimodal densities whose support include all the unimodal densities on the real line. Our motivation for this is that a mixture of the proposed univariate model can be adequate to cluster data, with each cluster being represented by a unimodal density.

One of our objectives was also to be able to create a class of unimodal densities that could be naturally extended to the multivariate case. Our models allow this through the use of the multivariate Gaussian copula. This way, modeling multivariate clusters can also benefit from the methodology developed in this paper.
  
The proposed models, however, cannot be dealt with analytically. We have also proposed an MCMC algorithm to obtain samples of the posterior distribution of the model parameters and to obtain predictive densities. The methodology was illustrated with univariate and bivariate examples, and with two variables taken from the Boston Housing Data (\cite{harrison78}). 

Modeling multivariate unimodal densities with dimension higher than two can be easily done with a slight modification to the code presented in section \ref{bivariate}. In that case, instead of sampling from a single correlation parameter $\rho$, we must sample from a correlation matrix, which can be done following \cite{Wu14}. Preliminary results showed this to be effective for a three-variate model.
For future work we must test the code extensively for three or more dimensions, and finally do clustering for an arbitrary dimension $d$, through the mixture of the densities proposed in this paper. 

\section{Acknowledgements*}
The first author acknowledges the support of a research grant from CNPq-Brazil.

\pagebreak

\end{document}